 \documentstyle[aps,multicol,psfig,eqsecnum]{revtex}  
\input{epsf}
\begin{document}
\def\tilt{{\chi}}
\def\newt{{X}}
\def\ss{\textstyle}
\draft
\title{Antiferromagnetic hedgehogs with superconducting cores}
\author{Paul M.~Goldbart\cite{REF:PMG} 
    and Daniel E.~Sheehy\cite{REF:DES}}
\address{
Department of Physics and Materials Research Laboratory, \\
University of Illinois at Urbana-Champaign, 
Urbana, Illinois 61801, USA}
 \date{P-97-10-030-iii; April 7, 1998}
\maketitle
\begin{abstract}
Excitations of the antiferromagnetic state that resemble
antiferromagnetic hedgehogs at large distances but are predominantly
superconducting inside a core region are discussed within the context of
Zhang's SO(5)-symmetry--based approach to the physics of
high-temperature superconducting materials.  Nonsingular, in contrast
with their hedgehog cousins in pure antiferromagnetism, these 
texture excitations are what hedgehogs become when the antiferromagnetic order
parameter is permitted to \lq\lq escape\rq\rq\ into superconducting
directions.  The structure of such excitations is determined in a simple
setting, and a number of their experimental implications are examined.
\end{abstract}
\pacs{74.20.De, 74.25.Dw, 74.25.Ha, 75.50.Ee}
%
%
%
	\begin{multicols}{2}
\narrowtext
\section{Introduction}
\label{SEC:intro}
In Zhang's SO(5)-symmetry--based approach to the physics of
high-temperature superconducting materials~\cite{REF:Zhang}, the local
state of the system at the spatial point ${\bf r}$ is characterized by
the orientation of a five-component unit vector ${\bf n}({\bf r})$.
Orientations for which $\sum_{a=1}^{3}\left(n^{a}\right)^{2}=0$ are
purely superconducting, the orientation of ${\bf n}$ in the
(two-dimensional) $4$-$5$ hyperplane determining the phase of the complex
superconducting order parameter. Orientations for which
$\sum_{a=4}^{5}\left(n^{a}\right)^{2}=0$ are purely antiferromagnetic,
the orientation of ${\bf n}$ in the (three-dimensional) $3$-$4$-$5$
hyperplane determining the direction in real space of the
antiferromagnetic (i.e.,\ N\'eel) vector order parameter. The novelty of
Zhang's approach lies in its assembling of these two order parameters
into a unified order parameter ${\bf n}$, and the consequent possibility
of orientations of ${\bf n}$ that do not lie wholely in one or other of
the superconducting and antiferromagnetic subspaces, instead
simultaneously containing components from both subspaces and, hence,
characterizing regions that are at once partially superconducting and
partially antiferromagnetic.  The purpose of the present Paper is to
point out a simple but potentially interesting property of this model:
in antiferromagnetic  regions of the phase diagram this model supports
three-(spatial)-dimensional antiferromagnetic hedgehog configurations
that find it energetically favorable to have superconducting cores, as 
depicted schematically in Fig.~\ref{FIG:hedge}. 

It should  be noted that the subject of the present Paper is, loosely
speaking, conjugate to that of the recent one~\cite{REF:Arovas} in which 
it is shown that, within the SO(5) approach, the cores of vortices in the 
superconducting order parameter should not be singular, the mechanism for the 
evasion of a singularity being escape from the two superconducting dimensions
into the three antiferromagnetic ones.

\section{Antiferromagnetic hedgehogs with superconducting cores}
\label{SEC:hedge}

What we mean by antiferromagnetic hedgehog configurations with 
superconducting cores are energetically-stationary spatial configurations 
of the order parameter ${\bf n}({\bf r})$ having the following properties:
(i)~Far from the (arbitrarily-located) center, the configuration 
${\bf n}({\bf r})$
closely resembles a (nonsuperconducting) 
antiferromagnetic hedgehog (i.e.,\ a point defect in which 
the N\'eel vector points radially away from the center, or some 
global SO(3) rotation of this configuration) and   the quantity 
$\sin^{2}\!\tilt$  $\big[\equiv\sum_{a=4}^{5}\left(n^{a}\right)^{2}\big]$, 
which measures the degree of superconducting order (without regard to 
its phase), is small.  Correspondingly, the complement 
$\cos^{2}\!\tilt$ $\big[\equiv\sum_{a=1}^{3}\left(n^{a}\right)^{2}\big]$, 
which measures the degree of antiferromagnetic order without 
regard to its orientation, is close to unity. 
(ii)~As the center of the configuration is approached, however, the
order parameter escapes from dimensions 1, 2 and 3 into
dimensions 4 and 5, so that superconducting order is acquired at the
expense of antiferromagnetic order.  Said equivalently, the angle
$\tilt$ rotates from $0$ to $\pi/2$ as the center of the configuration
is approached.  (In principle, more exotic hedgehog excitations are 
possible, in which the antiferromagnetic order varies more rapidly.
For the sake of simplicity we shall primarily focus on the simplest class.) 
By this mechanism, the medium is able to remain nonsingular, and
evade the (albeit finite) free-energy cost of the spatial
gradient in the N\'eel vector (this gradient diverging as the center of
the singular, purely antiferromagnetic, configuration is approached) at
the expense of condensing locally into the \lq\lq wrong\rq\rq\ (i.e.,
superconducting) state. 
(iii)~Whilst not being stable global-energetically---the homogeneous
antiferromagnetic configuration of course having a lower free
energy---antiferromagnetic hedgehog configurations with superconducting
cores do turn out to be energetically favorable,
compared with purely antiferromagnetic hedgehogs, as we shall see below,
at least when amplitude variations of the order parameter are inhibited.  
Presumably, such configurations are also {\it locally\/} energetically stable~\cite{REF:stable}. From the physical perspective, then, it would 
be quite intriguing if 
local regions of superconductivity were created by \lq\lq stressing\rq\rq\ 
the antiferromagnetism in regions of the phase diagram in which the stable 
homogeneous state is not superconducting.  Moreover, the topological 
stability of these textures will tend to hold these \lq\lq stresses\rq\rq\ 
in place.
 \begin{figure}[hbt]
 \epsfxsize=3.0truein
 \vskip-1.2truecm
 \centerline{\epsfbox{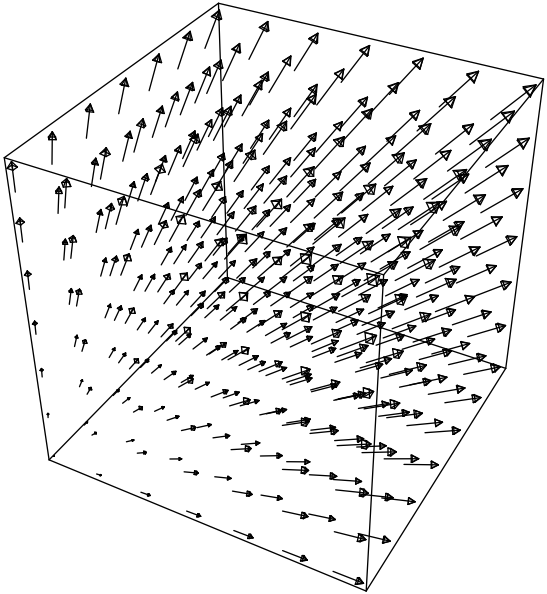}} 
 \vskip-1.0truecm
 \caption{An eighth of an antiferromagnetic hedgehog having a
superconducting core (determined numerically). The local orientation of
the vectors indicates the local orientation of the antiferromagnetism.
Their local magnitude indicates the local strength of the
antiferromagnetism and, hence, the local weakness of the
superconductivity.}
 \label{FIG:hedge}
 \end{figure}

In the simplest version of Zhang's approach, the free energy $F$ of a
three-dimensional sample in which the order parameter ${\bf n}({\bf r})$
varies with position ${\bf r}$ is given by
\begin{equation}
F=\int d^{3}r\,
\left\{
{\rho\over{2}}\sum_{\nu=1}^{3}\sum_{a=1}^{5}
\left(\partial_{\nu}\,n^{a}\right)^{2}
+{g\over{2}}\sum_{a=4}^{5}\left(n^{a}\right)^{2}
\right\}, 
\label{EQ:free}
\end{equation} 
where $\rho$ is the appropriate stiffness, $\nu$ ($=1,2,3$) runs through 
the cartesian spatial coordinates, and spatial anisotropies in the gradient 
term have been accommodated by coordinate rescalings.  By tuning the 
chemical potential $\mu$ relative to its critical value $\mu_{\rm c}$
(e.g., by doping), the parameter $g$ 
[$\propto\left(\mu_{\rm c}^2-\mu^2\right)$] is varied such that one moves
from a region in which antiferromagnetic states are favored ($g>0$)
to   a region in which superconducting   states are favored ($g<0$). 
This free energy is invariant under separate rotations on the
three-dimensional antiferromagnetic and two-dimensional superconducting
subspaces; invariance under arbitrary five-dimensional rotations is absent
whenever $g\ne 0$. Thus, from any configuration ${\bf n}({\bf r})$ one
can obtain a configuration having the same free energy via the
transformation
\begin{equation}
{\bf n}\rightarrow\left(R^{\rm A}\oplus R^{\rm S}\right)\cdot{\bf n}, 
\end{equation}
where $R^{\rm A}$ is a ($3\times 3$) orthogonal matrix operative in the
antiferromagnetic (i.e.,\ $a=1,2,3$) sector (i.e.,\ a magnetization
rotation) and $R^{\rm S}$ is a ($2\times 2$) orthogonal matrix operative
in the superconducting (i.e.,\ $a=4,5$) sector (i.e.,\ a phase rotation),
and the symbol $\oplus$ indicates that the five-dimensional operator is
block-diagonal and composed of one three- and one two-dimensional block.

To calculate the structure of an isolated antiferromagnetic hedgehog
with a superconducting core, let us make the hypothesis that components
of ${\bf n}({\bf r})$ in this configuration can be expressed in the form
\begin{equation}
 \pmatrix{
n^{1}\cr n^{2}\cr n^{3}\cr n^{4}\cr n^{5}}
=\pmatrix{
\cos\tilt(r)\,\sin\theta\,\cos\phi\cr 
\cos\tilt(r)\,\sin\theta\,\sin\phi\cr 
\cos\tilt(r)\,\cos\theta	  \cr
\sin\tilt(r)			  \cr 
0				     }.
\label{EQ:hypothesis}
\end{equation}
Here, $r$, $\theta$ and $\phi$ are spherical polar spatial 
coordinates centered
on the center of the configuration, and the function $\tilt(r)$, which
allows for interpolation between purely antiferromagnetic and purely
superconducting values of the order parameter, is assumed to depend only
on the radial distance from the center.
This configuration is spherically
symmetric, in the sense that
for it we have
\begin{equation}
{\bf n}(R^{\rm A}\cdot{\bf r})=
\left(R^{\rm A}\oplus I^{\rm S}\right)\cdot{\bf n}({\bf r}),
\end{equation}
where $I^{\rm S}$ is the identity operation in the superconducting sector. 
By exchanging the radial variable $r$ for the dimensionless version 
$t$ (i.e.,\ the radius, measured in units of the correlation length 
$\xi_{\pi}\equiv \sqrt{\rho/g}$ for 
the conversion of antiferromagnetic order into superconducting order) via
\begin{mathletters}
\begin{eqnarray}
\tilt(r)&\equiv&\newt(t),\\
r       &\equiv&\sqrt{\rho/g}\,t,
\end{eqnarray}
\end{mathletters}%
and inserting the configuration~(\ref{EQ:hypothesis}) into the free 
energy~(\ref{EQ:free}), we find that the free energy is given by
\begin{equation}
F=\tilde{F}\int_{0}^{\tau}dt\,
\left\{
 {t^{2}\over{2}}\dot{\newt}(t)^{2}
+\cos^{2}\newt(t)
+{t^{2}\over{2}}\sin^{2}\newt(t)
\right\}, 
\end{equation}
where $\tilde{F}\equiv 4\pi g(\rho/g)^{3/2}$, the overdot denotes a
derivative with respect to $t$,  and $\sqrt{\rho/g}\,\tau$ is a
large-distance cutoff, introduced to render finite the otherwise
linearly-divergent free energy.  Application of the calculus of
variations to the functional $F$ then leads to the stationarity 
condition
\begin{equation}
t^{2}\ddot{\newt}+2t\dot{\newt}
	+\left(1-{t^{2}\over{2}}\right)\sin 2\newt=0.  
\label{EQ:station}
\end{equation} 
 \begin{figure}[hbt]
 \epsfxsize=3.5truein
 \vskip-1.0truecm
 \centerline{\epsfbox{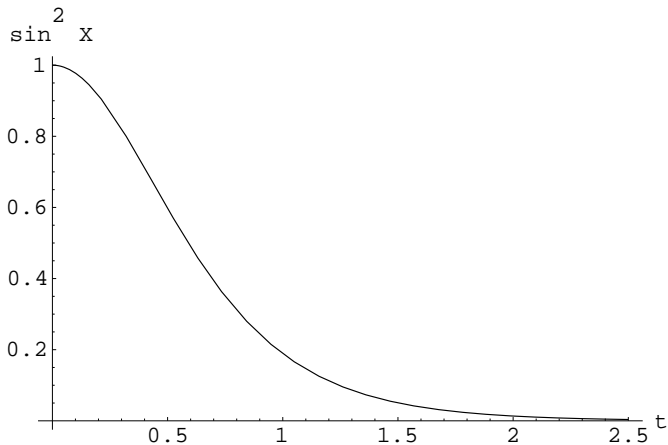}}
\caption{The degree of superconductivity $\sin^{2}\newt$ as a 
function of the scaled radial coordinate $t$ (determined numerically).
\label{FIG:numeric}}
\end{figure}
The relevant solutions of Eq.~(\ref{EQ:station}) are: 
(i)~$\newt(t)\equiv 0$ 
(i.e.,~\ the pure antiferromagnetic hedgehog, unescaped into the 
superconducting directions); and 
(ii)~the solution in which $\newt(t)$ interpolates between $\pi/2$ and $0$ 
as $t$ varies from $0$ to $\infty$ (i.e.,~\ the antiferromagnetic 
hedgehog with a superconducting core).  The precise form of the latter 
solution is readily found numerically, and is shown in 
Fig.~\ref{FIG:numeric}.  Its asymptotic behavior is 
$\big({\scriptstyle 1\over{\scriptstyle 2}}\pi-\newt\big)\sim t$ 
(for $t\ll 1$) and 
$\newt\sim\exp(-t)$ (for $t\rightarrow\infty$).
The configuration corresponding to solution (ii) is 
depicted in  Fig.~\ref{FIG:hedge}.  

To determine which of the solutions, (i) or (ii), has the lower free 
energy, let us consider the quantity 
$\Delta F\equiv\left(F^{\rm (ii)}-F^{\rm (i)}\right)$, 
where $F^{\rm (i)}$ and $F^{\rm (ii)}$ respectively refer to the 
free energy of solution~(i) and of solution~(ii). Then, by using 
Eq.~(\ref{EQ:station}), along with integration by parts,  we find that  
\begin{equation}
\Delta F=\tilde{F}\int_{0}^{\infty}dt\,
\left\{{t^{2}\over{2}}-1\right\}
\Big\{-\newt\cos\newt + \sin\newt\Big\}
\sin\newt, 
\label{EQ:increase}
\end{equation} 
where convergence at large $t$ permits the replacement of the upper 
limit by $\infty$. The numerical evaluation of this quantity gives 
$\Delta F\approx -0.272\,\tilde{F}$.  This indicates that it is 
energetically favorable for the order parameter in the core of an 
antiferromagnetic hedgehog to escape into the superconducting 
directions.
\section{Consequences of amplitude-sector fluctuations}
\label{SEC:amplitude}
As we have seen, in the setting of a model in which the amplitude of
${\bf n}({\bf r})$ is constrained to be unity, there are hedgehog
excitations that have superconducting cores.  We now explore the issue
of whether such excitations continue to exist in settings in which
amplitude variations of ${\bf n}({\bf r})$ are inhibited (i.e.,\ 
are not prohibited, although
they are suppressed energetically). Under such circumstances, it is
possible---and may prove energetically favorable---for the core of the
hedgehog to avoid antiferromagnetic gradient energy via the 
development of an amplitude-reduced purely-antiferromagnetic core, 
rather than by escaping into the superconducting directions.
To address this issue, we follow Arovas et al.~\cite{REF:Arovas} and 
consider a \lq\lq soft-spin\rq\rq\ generalization of the SO(5) model. 
Thus we consider a free energy of the form 
\begin{eqnarray}
F&=& 
\frac{\ss\rho}{\ss 2}
\int d^3r\,
\left\{
 \left(\partial_{r}n(r)\right)^2 
+2r^{-2}\,n(r)^2\,\cos^{2}\!\chi(r) 
\right.
\nonumber
\\&&\qquad
\left.
+\,n(r)^2\,\left(\partial_r\chi(r)\right)^2 
+\xi_{\pi}^{-2}\,n(r)^2\,\sin^2\!\chi(r) 
\right\}
\nonumber
\\&&\qquad\qquad
+\,a\int d^3r\,
\Big\{-\frac{\ss 1}{\ss 2}n(r)^2 + \frac{\ss 1}{\ss 4}n(r)^4\Big\}, 
\label{EQ:amplitude} 
\end{eqnarray}
where $2a$ denotes the squared \lq\lq mass\rq\rq\ associated with the
amplitude-sector fluctuations of ${\bf n}$, $n$ is the amplitude
of ${\bf n}$, and we have restricted the discussion to (spatially)
spherically symmetric configurations. 

In the absence of amplitude fluctuations, 
the hedgehog with superconducting core has  $n \equiv 1$ and $\chi $  
varying from $0$ to $\pi/2$ as the center of the texture is approached.  
The amplitude-reduced purely antiferromagnetic 
hedgehog excitation  will be one for which  $n$ 
vanishes at 
the center of the texture
and grows to unity at large distances, and
$\chi\equiv 0$.  The stationarity condition for  $n$, which determines 
the structure of the amplitude-reduced hedgehog, can be solved
numerically, allowing us to obtain the function $n(r)$.  Then, we
may insert this back into Eq.~(\ref{EQ:amplitude}) to obtain the 
free energy of the purely antiferromagnetic
hedgehog with reduced-amplitude core. 
The free energies of the hedgehog with superconducting
core and the purely antiferromagnetic
hedgehog with amplitude-reduced core each must be defined
with a long distance cutoff to render them finite, but the difference 
between these two quantities is independent of this cutoff, and
turns out to be  given by
\begin{equation}
F_{\rm AF}-F_{\rm SC}\approx
4\pi\rho\left(
  0.272\sqrt{\rho/g} 
- 1.454\sqrt{\rho/a}
\right),
\end{equation}
where the subscripts refer to the purely-anti{\-}ferro{\-}magnetic 
amplitude-reduced (AF) and superconducting-core (SC) hedgehogs.
Thus, within the context of this model in which amplitude sector 
fluctuations are permitted, we find that the hedgehog with 
superconducting core will be energetically preferred when this 
quantity is positive, i.e., provided that $g<0.035 a$ 
[or, equivalently, $\xi_{\pi}>5.35\xi_{\rm a}$, 
where $\xi_{\rm a}$ ($\equiv\sqrt{\rho/a}$) denotes the fluctuation 
correlation length for antiferromagnetic fluctuations].
Now, it is typical for $\xi_{\rm a}$ to be on the order of a lattice
spacing for the cuprate materials, whereas $\xi_{\pi}$ is expected to
grow as the superconducting phase boundary is approached from the
antiferromagnetic state.  Thus,
one should anticipate that over a substantial portion of the 
antiferromagnetic part of the phase diagram, antiferromagnetic
hedgehog excitations will have escaped-superconducting (rather than
amplitude-reduced purely antiferromagnetic) cores.
\section{Topological classification of hedgehog excitations}
\label{SEC:topology}
We now turn to the issue of the topological classification of hedgehog
excitations having superconducting cores, these excitations being
nonsingular textures of the order-parameter field 
${\bf n}({\bf r})$.  In pure antiferromagnets, the existence of singular,
hedgehog point-defect excitations is expressed, mathematically, by the
statement $\Pi_2(S_2) = Z$~\cite{REF:EDM,REF:Mermin}. What this means is
that mappings (provided by order-parameter configurations) of spheres in
real-space into the antiferromagnetic order-parameter space $S_2$ fall
into homotopically inequivalent classes labeled by the integers (and 
combine according to integer arithmetic).  Within the SO(5) approach,
however, the nonsingular hedgehog texture excitations having
superconducting cores are described by order parameter configurations
${\bf n}({\bf r})$ that lie in the antiferromagnetic subspace $S_2$ at
large distances from the core, but escape into the full order-parameter 
space $S_4$, as the core is approached.  In order to complete the 
classification of these textures, then, we should ascertain whether or 
not there exist homotopically inequivalent textures that, at large 
distances, are homotopically equivalent and lie in the antiferromagnetic 
subspace $S_2$.  The appropriate mathematical machinery for this task 
involves {\it relative homotopy groups\/} and {\it exact homotopy 
sequences\/}~\cite{REF:Mermin,REF:BL}.  
To implement this machinery, we consider 
mappings of cubes (in real-space) such that the surface of the cube is 
mapped into $S_2$ whereas the interior of the cube is mapped into $S_4$.  
Such mappings are classified according to the relative homotopy group 
$\Pi_3(S_4, S_2)$.  This group is readily computed by making use of the 
exact sequence of homomorphisms:
\begin{equation}
\Pi_3(S_4)
\stackrel{\beta_3} \rightarrow\Pi_3(S_4, S_2) 
\stackrel{\gamma_3}\rightarrow\Pi_2(S_2) 
\stackrel{\alpha_2}\rightarrow\Pi_2(S_4).
\end{equation}
Here, $\beta_3$, $\gamma_3$ and $\alpha_2$ denote mappings of the 
elements of the previous group in the sequence to elements of the 
following group, that, in general, are not isomorphic~\cite{REF:footnote}.  
Now, as $\Pi_3(S_4)$ and $ \Pi_2(S_4) $ are both the trivial group, the 
homomorphism $\gamma_3$ is, in fact, an isomorphism~\cite{REF:Mermin,REF:BL}, 
from which it follows that $\Pi_3(S_4, S_2)\cong\Pi_2(S_2)\cong Z$ and, 
thus, we find that there is no structure in $\Pi_3(S_4, S_2)$ beyond what 
was already present in $ \Pi_2(S_2)$.  
The physical consequence of this result is that whilst hedgehog excitations
fall into homotopically inequivalent classes, the possible nonsingular superconducting cores of a given class of hedgehog are homotopically
equivalent to one another. 
\section{Related structures in other condensed states}
\label{SEC:relate}
The notion of the conversion of singularities into textures via the 
escaping of order-parameters into additional directions has been realized 
in several other condensed matter settings.  For example, (uniaxially) 
nematic liquid-crystalline media
have long been known to exhibit a structure closely related to
antiferromagnetic hedgehog configurations with superconducting cores.
When confined to a cylinder that imposes homeotropic (i.e., perpendicular)
boundary conditions on the nematic alignment, the system can evade the
threading of the cylinder by a singularity because the order parameter
orientation can escape from the radial plane into the
axial direction~\cite{REF:escape}.  This mechanism remains energetically
favorable even for diamagnetic nematics in an axial magnetic field (for which
escape costs condensation energy).

Superfluid $^3$He is another system that provides a rich array of 
topologically interesting textures~\cite{REF:relations}.  The example
having the most relevance to the present Paper is that of hedgehog 
excitations in  $^3$He-B.  The order parameter for $^3$He-B is a 
complex-valued $3\times 3$ matrix of the form 
$A_{\mu \nu}=e^{i\phi}R_{\mu\nu}({\bf \hat{n}},\theta)$, where 
$\phi$ is a phase angle and $R_{\mu\nu}$ is a rotation matrix about the 
unit vector ${\hat{\bf n}}$ by an angle $\theta$.  On long length-scales, 
$\theta$ becomes fixed, due to a dipolar coupling, acquiring the value
$\theta_{\rm L}$, known as the Leggett angle, so that the low-energy degrees 
of freedom are expressed by the possible values of $\phi$ and the directions 
of $\hat{{\bf n}}$.  Thus, the order-parameter space ${\rm G}$ is given by 
${\rm G}={\rm U}(1)\times S_2$, so that 
$\Pi_2({\rm G})=\Pi_2(S_2)=Z$, so that the system may form hedgehogs with 
the unit vector $\hat{\bf n}$.  On short length-scales, however,  
$\theta$ can vary, so that the order-parameter space is effectively enlarged 
to ${\rm U}(1)\times {\rm SO}(3)$.  We see that, as 
$\Pi_2\big({\rm SO}(3)\big)=0$, over short distances there are no
topologically stable point-defects, so that hedgehogs in  $^3$He-B have 
nonsingular cores.  

A similar effect occurs in nematic liquid crystals~\cite{REF:Lyuksyutov}, 
where on large length-scales the system is uniaxial, so that the relevant 
order-parameter space is $RP_2$ (i.e., the real projective plane constructed 
by identifying opposite points on $S_2$).  On short length-scales, however, 
the order-parameter space is enlarged to $S_4$, so that disclinations in 
the nematic order have nonsingular cores. 

\section{Experimental signatures of hedgehogs; concluding remarks}
\label{SEC:experiment}
We now briefly consider some issues associated with antiferromagnetic
hedgehogs having superconducting cores that might be relevant to experiments.
These excitations should be present after performing a quench, from high
temperature or high magnetic field, into the antiferromagnetic state.
The number of excitations per unit volume should be higher for
more rapid quenches.  As time proceeds after the quench, the number of
hedgehog excitations can decrease via their mutual annihilation,
although this would require collisions of two or more hedgehogs.
Presumably, this process can occur relatively slowly, at least at
sufficiently low temperatures, so that one may anticipate regimes in
which these excitations, once created,  remain long enough for their
consequences to be detected.

What experimental signatures might antiferromagnetic hedgehogs with
superconducting cores yield? Let us suppose that a sufficiently high 
density of such excitations can be created, and that this density 
can be maintained for a sufficiently long time.  Then one may 
crudely regard the excitations as providing a set of randomly located, 
randomly phased, superconducting inclusions~\cite{REF:coherent}.  
These inclusions would not be unlike Aslamazov-Larkin paraconducting
fluctuations~\cite{REF:fluctuate}, except that  they would be  
\lq\lq externally\rq\rq\ maintained and, therefore, could be much 
longer lived.  One might hope that these inclusions would be detectable 
in electrical conductivity experiments, their presence leading to an
enhancement of the conductivity.  (One would need to account for
scattering from the antiferromagnetic hedgehogs which, presumably,
diminishes the conductivity.)\thinspace\  This enhancement should be 
suppressed by magnetic fields, and by the decay of the excitations. 
Similarly, one might also envisage observing Andreev reflection from 
the superconducting inclusions (although capacitive charging effects
may suppress this effect\cite{REF:BA}).  

An externally applied magnetic field will be partially screened by these 
inclusions, leading to a negative contribution to the magnetic susceptibility. 
To estimate the size of this effect, we approximate the hedgehog cores 
to be uniformly superconducting and spherical in shape~\cite{REF:Tinkham}.  
In the regime where the London penetration depth $\lambda$
is much longer than the core radius $ \xi$, this leads to a 
diamagnetic susceptibility contribution 
$\chi=-\delta\xi^5/40\pi\lambda^2$,
where $\delta $ is the number of excitations per unit volume.
One might also hope that the presence of antiferromagnetic hedgehogs
with superconducting cores would be detectable via probes such as
nuclear magnetic resonance, electromagnetic absorption
and hedgehog/antihedgehog pair creation and, perhaps fancifully, 
scanning tunneling  microscopy (e.g.,\ with a magnetic tip). 

In addition, 
these excitations should leave their fingerprint on the (staggered) 
magnetic structure factor $S(k)$, this factor being determined by 
${\bf N}({\bf k})$ (i.e., the Fourier transform of the antiferromagnetic 
N\'eel vector at the probing wave vector ${\bf k}$) via
\begin{equation}
S(k)\equiv V^{-1}\,{\bf N}({\bf k})\cdot{\bf N}({\bf k}), 
\end{equation}
where $V$ is the volume of the system. 
Specifically, for length scales that are long compared with the core 
size $\xi_{\pi}$ but short compared to the spacing between the 
hedgehogs $\delta^{-1/3}$, we expect $S(k)$ to have the conventional 
hedgehog form.  However, for length scales that are short compared 
with the core size but long compared with the lattice spacing, 
we expect a reduction in $S(k)$ (and hence scattering), 
owing to the diminution of the Fourier amplitude of the 
antiferromagnetic moment at these scales.  The computed 
hedgehog structure does indeed realize this scenario: 
\begin{equation}
S(k)\sim
\cases{
\delta\,\xi^{6}\,(k\xi)^{-6}
	&for $\delta^{-1/3}\ll k\ll\xi_{\pi}^{-1}$;\cr 
\delta\,\xi^{6}\,(k\xi)^{-10}
	&for $\xi_{\pi}^{-1}\ll k$.\cr}
\end{equation}
However, it should be noted that, being sensitive only to 
antiferromagnetic order, this particular probe does not directly
ascertain whether or not the cores of hedgehog excitations are
superconducting (except via the dependence of the size of the cores on
the location of the system in the phase diagram).

If detected in experiments, antiferromagnetic hedgehogs with
superconducting cores would provide striking evidence in support of
Zhang's SO(5) approach to the physics of high-temperature
superconducting materials.  Their presence would corroborate the 
notion that superconducting excitations are essential low-energy
excitations of the antiferromagnetic state.  Moreover, it would
prove rather intriguing to have at hand a physical system in which
superconductivity is induced by the distortion of a thermodynamically
preferred nonsuperconducting state.

\smallskip
\noindent
{\it Acknowledgments\/} ---  
We acknowledge helpful discussions with  Yuli Lyanda-Geller, Kieran Mullen 
and Martin Z\'apotock\'y, and thank G.~Volovik for bringing to our attention
related phenomena in other condensed matter systems.
This work was supported by 
the U.S.~Department of Energy, Division of Materials Sciences, 
under Award No.~DEFG02-96ER45439 through the University of Illinois 
Materials Research Laboratory.

	\end{multicols}

\begin{references}
\bibitem[\ast]{REF:PMG}
E-mail address: {\tt goldbart@uiuc.edu\/}
\bibitem[\dagger]{REF:DES}
E-mail address: {\tt d-sheehy@uiuc.edu\/}
\bibitem{REF:Zhang}
S.-C.\ Zhang, 
Science {\bf 275\/}, 1089 (1997). 
\bibitem{REF:Arovas}
D.\ P.\ Arovas, A.\ J.\ Berlinsky, C.\ Kallin and S.-C.\ Zhang, 
Phys.\ Rev.\ Lett.\ {\bf 79}, 2871 (1997).
\bibitem{REF:stable}
Stability with respect to small variations has not yet been checked 
for the stationary antiferromagnetic hedgehog configuration with a 
superconducting core. 
\bibitem{REF:EDM} 
For a table of homotopy groups see the 
{\sl Encyclopedic Dictionary of Mathematics\/}, 
edited by Kiyosi It\^o (MIT Press, Cambridge, MA, 1993), 
App.~A, Table~6.VI.
\bibitem{REF:Mermin}
For pedagogical introductions to the topological theory of defects in
condensed matter, see:
N.\ D.\ Mermin, 
Rev.\ Mod.\ Phys.\ {\bf 51\/} 591 (1979); and  
M.\ Kl\'eman, 
{\sl Points, Lines, and Walls: in liquid crystals, magnetic systems, 
and various ordered media\/} (John Wiley, New York, 1983)
\bibitem{REF:BL}
For application of this machinery in the context of $^3$He, see 
D. Bailin and A. Love, 
J. Phys. A: Math. Gen.~{\bf 11\/},  821 (1978);  
J. Phys. A: Math. Gen.~{\bf 11\/}, L219 (1978). 
For a general discussion of homotopy theory, see 
N. Steenrod, 
{\sl The Topology of Fibre Bundles\/} 
(Princeton University, Princeton, NJ, 1951).
\bibitem{REF:footnote}
This means that the homomorphisms are not necessarily {\it onto\/}
(i.e., do not  map onto every element of the following group) and are
not necessarily {\it one-to-one\/} (i.e., some elements of one group may
be mapped into more than one element of the following group).
\bibitem{REF:escape}
P.\ E.\ Cladis and M.\ Kl\'eman, 
J.\ Phys.\ (Paris) {\bf 33\/}, 591 (1972); 
R.\ B.\ Meyer, 
Phil.\ Mag.\ {\bf 27\/}, 405 (1973); 
for a discussion, see 
P.\ G.\ De~Gennes and J.\ Prost, 
{\sl The Physics of Liquid Crystals\/} 
(Oxford University Press, 1993), Sec.~4.3.1. 
\bibitem{REF:relations}
G. E. Volovik and V. P. Mineyev, Zh. Eksp. Teor. Fiz.
Pis'ma Red. {\bf 24}, 605 (1976) [JETP Lett. {\bf 24}, 561 (1976)].
For a discussion of textures in $^3$He, see  Chap.~7 
of D.~Vollhardt and P.~Wolfle,
    {\sl Superfluid Phases of Helium-three\/}
    (Taylor and Francis, London, 1990), and references therein.
%
\bibitem{REF:Lyuksyutov}
I. F. Lyuksyutov, Zh. Eksp. Teor. Fiz. {\bf 75}, 358 (1976)
[JETP~{\bf 48\/} 178 (1978)]
\bibitem{REF:coherent}
It may be interesting to consider the implications of phase-coherence
between superconducting cores, although this may not be straightforward
to produce.
\bibitem{REF:fluctuate}
L.\ G.\ Aslamazov and A.\ I.\ Larkin, 
Phys.\ Lett.\ {\bf 26A\/}, 138 (1968);  
Fiz.\ Tverd.\ Tela {\bf 10\/}, 1104 (1968) 
[Sov.\ Phys.\ Solid St.\ {\bf 13\/}, 2474 (1972)].
For discussions, see 
I.\ O.\ Kulik and I.\ K.\ Yanson,  
{\sl The Josephson Effect in Superconductive Tunneling Structures\/} 
(Israel Program for Scientific Translation, Jerusalem, 1972);
M.\ Tinkham, 
{\sl Introduction to Superconductivity\/}
(McGraw-Hill, New York, 1975); Chap.~7; 
A.\ A.\ Abrikosov, 
{\sl Fundamentals of the Theory of Metals\/} 
(North-Holland, 1988), Sec.~19.6
\bibitem{REF:BA}
We thank Boris Altshuler for alerting us to the issue of charging effects.
\bibitem{REF:Tinkham}
For a discussion see Chap.~2 of the book by 
Tinkham cited in Ref.~\cite{REF:fluctuate}. 
\end{references}
\end{document}